\documentclass[twocolumn,amsmath,amssymb,prb,longbibliography]{revtex4-2}
\usepackage{graphicx}
\usepackage{dcolumn}
\usepackage{bm}
\usepackage{color}  

\begin{document} 



\title{A photonic platform hosting telecom photon emitters in silicon }

\author{Michael Hollenbach$^{1,2,5}$}
\author{Nagesh S. Jagtap$^{1,2,5}$}
\author{Ciar{\' a}n Fowley$^{1}$}
\author{Juan Baratech$^{1}$}
\author{Ver{\' o}nica Guardia-Arce$^{1}$} 
\author{Ulrich Kentsch$^{1}$}
\author{Anna Eichler-Volf$^{1}$}
\author{Nikolay V. Abrosimov$^{1}$}
\author{Artur Erbe$^{1}$}
\author{ChaeHo Shin$^{4}$}
\author{Hakseong Kim$^{4}$}
\author{Manfred Helm$^{1,2}$}
\author{Woo Lee$^{4}$}
\author{Georgy V. Astakhov$^{1}$}
\author{Yonder Berenc{\' e}n$^{1}$}
\email[E-mail:~]{y.berencen@hzdr.de}

\affiliation{Helmholtz-Zentrum Dresden-Rossendorf, Institute of Ion Beam Physics and Materials Research, 01328 Dresden, Germany  \\ 
Technical University of Dresden, 01062 Dresden, Germany  \\ 
Leibniz Institute for Crystal Growth, 12489 Berlin, Germany  \\  
Korea Research Institute of Standards and Science, 34113 Daejeon, Republic of Korea  \\ 
$^5$These authors contributed equally to this work}

\begin{abstract}  
Silicon, a ubiquitous material in modern computing, is an emerging platform for realizing a source of indistinguishable single-photons on demand. The integration of recently discovered single-photon emitters in silicon into photonic structures, is advantageous to exploit their full potential for integrated photonic quantum technologies. Here, we show the integration of telecom photon emitters in a photonic platform consisting of silicon nanopillars. We developed a CMOS-compatible nanofabrication method, enabling the production of thousands of individual nanopillars per square millimeter with state-of-the-art photonic-circuit pitch, all the while being free of fabrication-related radiation damage defects. We found a waveguiding effect of the 1278 nm-G center emission along individual pillars accompanied by improved brightness, photoluminescence signal-to-noise ratio and photon extraction efficiency compared to that of bulk silicon. These results unlock clear pathways to monolithically integrating single-photon emitters into a photonic platform at a scale that matches the required pitch of quantum photonic circuits.\\\\ 
Keywords: G-center, silicon nanopillars, photonic integration, metal-assisted chemical etching, ion implantation
\end{abstract}

\date{\today}

\maketitle

\section{Introduction}
The monolithic integration of a single-photon source with reconfigurable photonic elements and single-photon detectors in a single silicon chip would result in quantum photonic integrated circuits that would harness quantum phenomena for secure communication \cite{Wang2020} and computation \cite{Zhong2020}. These quantum photonic integrated circuits could be fabricated using the same equipment and methods as computer circuits, avoiding the high cost of developing a new technology. Yet, an on-demand source of indistinguishable single-photons in silicon is critically lacking.

Silicon has recently been proven to be instrumental for hosting sources of single-photons emitting in the strategic optical telecommunication O-band (1260-1360 nm) \cite{Hollenbach2020,Redjem2020,Durand2021}.  
The origin of the telecom single-photon emitters is a carbon-irradiation induced damage center in silicon, the so-called G-center \cite{Davies1989,Udvarhelyi2021}. 
These single-photon emitters are created at sufficiently low carbon concentrations, whereby two or more of the G center defects do not interact with each other \cite{Hollenbach2020}. 
The scalability, brightness, purity and collection efficiency of these single-photon emitters are key issues in bringing them closer to practical applications. 
A scalable photonic platform that houses single-photon emitters integrated into photonic structures such as nanopillars \cite{Radulaski2017}, solid immersion lenses \cite{Widmann2015} or high-quality optical microcavities \cite{Wachter2019} is thus beneficial. 

Reactive ion etching (RIE) is the method of choice in mainstream silicon integrated circuit manufacturing to transfer lithographically defined photoresist patterns into electronic and photonic materials with great fidelity \cite{Huff2021}. 
Yet, crystal damage along with contamination defects can be introduced during the RIE process of silicon since its surface is exposed to bombardment by energetic particles \cite{Huff2021}. 
Consequently, RIE of silicon also produces radiation damage defects with sharp luminescence lines, such as the X- (1192 nm), W- (1218 nm), G- (1278 nm), I- (1284 nm), T- (1326 nm), C- (1570 nm) and P-line (1616 nm)  \cite{Davies1989,Weber1986,Henry1991}. 
The uncontrollable creation of the type, position and density of these defects is not desirable for practical quantum applications.
Alternatively, metal-assisted chemical etching (MACEtch) is a promising etching technique for defect-free pattern transfer in silicon, which enables the fabrication of high-aspect ratio structures \cite{Han,Li2012} for applications in photovoltaics \cite{Li2012}, X-ray optics \cite{Romano2020}, energy storage \cite{Arico2005} and sensors \cite{Huang2007}. 
MACEtch is a wet-chemical process in which vertical etch fronts are obtained in silicon using only a thin-film noble metal mask (e.g. Au, Ag, Pt, Pd) and an etching solution consisting of a mixture of a strong oxidizing agent (e.g. H$_{2}$O$_{2}$) and hydrofluoric acid (HF). 
Si is oxidized in the region where it is in contact with the metal mask, which catalyzes the oxidation process by the injection of holes into the Si valence band. 
The HF subsequently dissolves the formed SiO$_{2}$. 

Here, we demonstrate the integration of 1278 nm emitting G-centers in a two-dimensional photonic platform of arrays of silicon nanopillars. 
Our CMOS-compatible nanofabrication approach enables the production of thousands of individual nanopillars per square millimeter each hosting a telecom photon emitter.
Moreover, the structures are free of fabrication-related radiation damage defects. 
Compared to that of bulk Si, we found an improved brightness, photoluminescence (PL) signal-to-noise ratio and photon extraction efficiency of the G center-related emission along with a waveguiding effect of the 1278 nm-photon emission along individual pillars. 
This effect is confirmed by the reduction by a factor of 3.5 of the full width at half maximum (FWHM) of the G-center emission along the pillar when compared to that of bulk Si.

\section{Experimental details}

\begin{figure}
  \centering
  \includegraphics[width=0.4\textwidth]{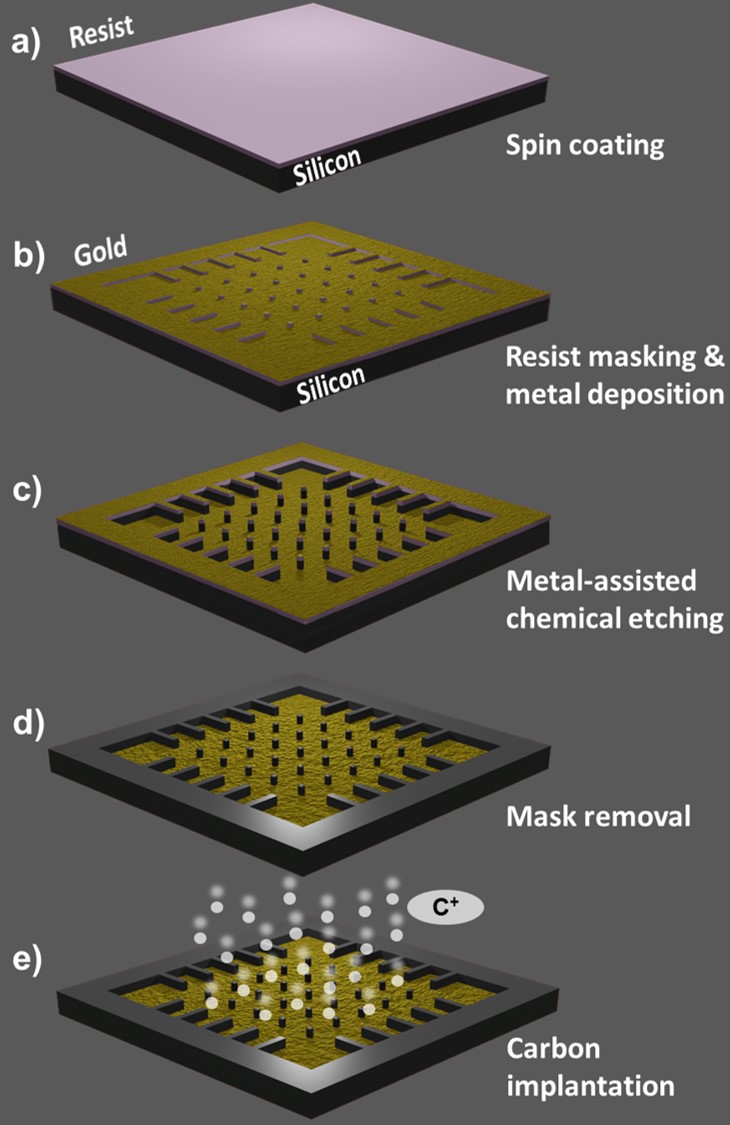}
  \caption{ Schematic of the process for the integration of telecom photon emitters into individual Si nanopillars. a) Spin coating of the photoresist. b) Pattern design of the two-dimensional array of nanopillars by electron beam lithography along with gold deposition. c) Metal-assisted chemical etching (MACEtch). d) photoresist-based mask removal using ultrasonification. e) Broad-beam carbon implantation with a mean projected range of carbon ions of around 600 nm in pillars with an average height of 1200 nm.}
  \label{processflow}
\end{figure}

We used $\langle 100 \rangle$-oriented single-side polished silicon substrates grown by float zone technique with an as-grown concentration of carbon and oxygen impurities less than 5$\times$10$^{14}$ cm$^{-3}$ and 1$\times$10$^{14}$ cm$^{-3}$, respectively. 
The $525 \, \mathrm{\mu m}$ thick substrates were cleaved into 5 mm x 5 mm pieces. Next, they were cleaned with Piranha solution (3 parts H$_{2}$SO$_{4}$ (96$\%$) : 1 part H$_{2}$O$_{2}$(31$\%$)) for 900 sec. 
The samples were then dipped in 0.5$\%$ HF solution to remove the native silicon oxide. A negative tone electron-beam resist was spin-coated (ma-N 2410, 3500 rpm, 45 sec) followed by post baking at 90$^{\circ}$ C for 150 sec (Fig. \ref{processflow} a).

The design pattern, consisting of two-dimensional nanopillar arrays with diameters ranging from 300 to 1100 nm, was transferred using electron beam lithography (EBL) tool (Raith 150TWO, acceleration voltage 20 kV, 20 $\mathrm{\mu}$m aperture with base dose 130  $\mathrm{\mu}$C cm$^{-2}$) followed by development in ma-D 525 for 210 sec (Fig. \ref{processflow} b). 
The 5 $\mathrm{\mu}$m pitch between all pillars was chosen for compatibility with standard photonic circuits \cite{Zhang2013}.
A 10 nm-thick gold was electron-beam evaporated at the rate of 3 \AA.sec$^{-1}$ after native oxide removal step with 0.5$\%$ HF dip step (Fig. \ref{processflow} b). 
MACEtch,  (Fig. \ref{processflow} c), was performed in a cleanroom environment at room temperature. 
We used a chemical solution consisting of 25 ml HF(40$\%$), 4.5 ml H$_{2}$O$_{2}$ (31$\%$) and 100 ml DI water. 
The sample was dipped in this solution facing up for 300 sec. 
The photoresist was subsequently removed (Fig. \ref{processflow} d) using ultrasonic agitation in acetone for 30 sec followed by washing in isopropyl alcohol and blow-drying under a stream of nitrogen gas. 
The sample was then cleaned by IPA. 
Without removing the gold thin film, we created the G center-based telecom photon emitters by implanting carbon at a fluence of 2$\times$10$^{14}$ cm$^{-2}$ with an energy of 250 keV. 
This implantation energy corresponds to a mean projected range of carbon ions within the pillars of around 600 nm (Fig. \ref{processflow} e).
A schematic of the process flow is shown in Figure \ref{processflow}.

The optical experiments were performed in a home-built low-temperature confocal photoluminescence microscope. A 637 nm-laser diode pigtailed with a single mode optical fiber is coupled to the confocal setup. The incident laser beam is reflected by a dichroic mirror and focused on the sample by a cryocompatible objective, which is optimized for the telecommunication wavelength range. The high-numerical aperture (NA=0.81) objective focuses the beam with a spot size of roughly 1 $\mathrm{\mu}$m on the sample surface with an estimated laser power of 3 mW. The silicon substrate, along with the fabricated arrays of different sized nanopillars, is mounted onto a sample holder inside a customized closed-cycle helium cryostat that ensures a stable sample temperature of 6 K. Using two linear nanopositioners anchored to the sample holder, two-dimensional PL mappings are performed. For the in-depth PL scans, the refractive index of silicon n$_{Si}$ = 3.48 is considered. The emitted PL from the G-center, is collected by the same objective before imaged onto a 500 $\mathrm{\mu}$m confocal pinhole. A set of long pass filters is used to fully suppress the above bandgap laser light and Si bandgap emission. The PL signal is sent to a broadband fiber-coupled superconducting nanowire single-photon detector. For PL spectra, the signal is fed into a spectrometer equipped with an InGaAs detector. Further details about the optical setup can be found elsewhere \cite{Hollenbach2020}.

\section{Results and Discussion}

\begin{figure}[h]
  \centering
  \includegraphics[width=0.4\textwidth]{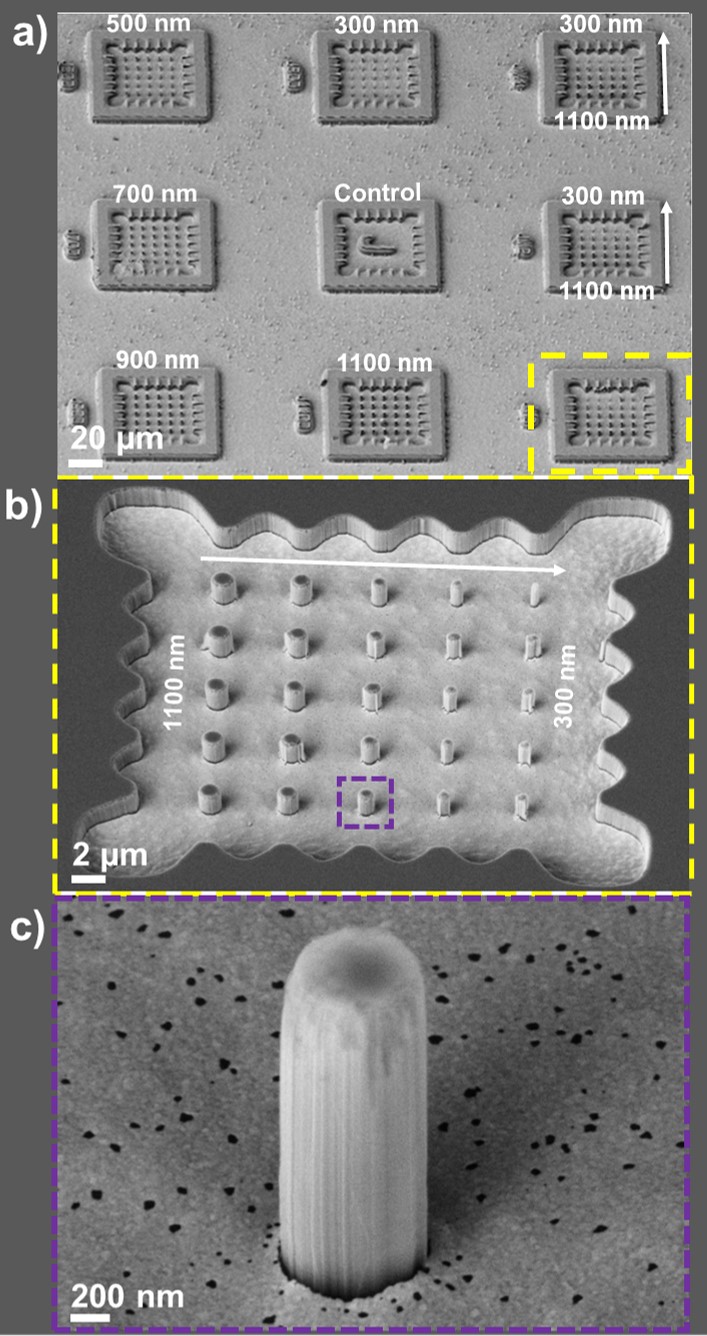}
  \caption{Scanning electron microscopy (SEM) images of MACEtched nanopillars taken at 45. a) 3 $\times$ 3 arrays of dies in a Si chip containing silicon nanopillars with different diameters fabricated by MACEtch. b) a 5 $\times$ 5 array of nanopillars with diameters of around 300, 500, 700, 900 and 1100 nm. The center structure serves as a control. c) An individual Si nanopillar with a diameter of 700 nm and a height of around 1200 nm.}
  \label{MACEtcharray}
\end{figure}

The two-dimensional pattern, see Fig. \ref{MACEtcharray} a, consists of 3 x 3 structures within an area of 250 $\mathrm{\mu}$m x 250 $\mathrm{\mu}$m. 
Each pattern containing 1 control array and 8 arrays of 5 x 5 nanopillars with diameters ranging from 300 to 1100 nm. 
Within each array (Fig. \ref{MACEtcharray} b), the center-to-center distance is set to 5 $\mathrm{\mu}$m to prevent crosstalk between adjacent individual pillars. 
Typically, a 3 to 5 $\mu$m pitch is used in integrated photonic circuits to avoid crosstalk losses between neighboring photonic elements in which the working wavelength is in one of the optical telecom bands \cite{Zhang2013}. 
Each fabricated pillar has a near vertical sidewall (Fig. \ref{MACEtcharray} c), which is an advantage of the MACEtch process. 
The pillar height is 1200 nm. 

There are mainly two processes in MACEtch for material transfer at the Si/metal interface. 
The reagent and byproducts diffuse along the periphery of the etched structure \cite{Peng2005,Peng2006} (Process I). 
The other mechanism is that Si atoms are dissolved and diffuse through the noble metal, are oxidized and then etched by HF \cite{Kobayashi1998,Xie2008,Buttner2008} (Process II). 
In our case, after MACEtching the 10 nm thick Au film clearly semi-porous, see Fig. \ref{MACEtcharray} c. 
As also seen in Fig. \ref{MACEtcharray} c, the diameter at the top of the pillar is slightly smaller than the rest.
Furthermore, a small up-turn of the Au film is also visible at the very bottom of the pillar. 
The damage to the Au film and the up-turn of the Au at the nanopillar periphery during MACEtching support the idea that Process I dominates.
The survival of the Au film should depend critically on the Au thickness if Process I predominates.
We find this is indeed the case with 30 nm Au films; the entire film is removed even after very short etching times.
Another key aspect of our design, which supports this conclusion, is the border-region around the nanopillar arrays.
We isolated each 5 $\times$ 5 array with a border region, 5 $\mathrm{\mu}$m in width, which separates the etched array from the extended Au film, allowing for more efficient diffusion of etched products away from the nanopillar structures and results in less film tearing. 
It can be seen in the SEM micrograph (Fig. \ref{MACEtcharray} a) that outside the frames, where the interframe distance is 50 $\mathrm{\mu}$m, numerous artifact structures are formed. 
Within the boundary of the frame (Fig. \ref{MACEtcharray} b), however, the etching uniformity and continuity of the gold film is superior. 
If Process II were to dominate we would not expect to see any difference inside and outside the array areas, nor would we expect a diameter increase change during etching.
An added advantage is that the border region also allows for easy location of the nanopillar arrays in the confocal microscope.

\begin{figure}
  \centering
  \includegraphics[width=0.38\textwidth]{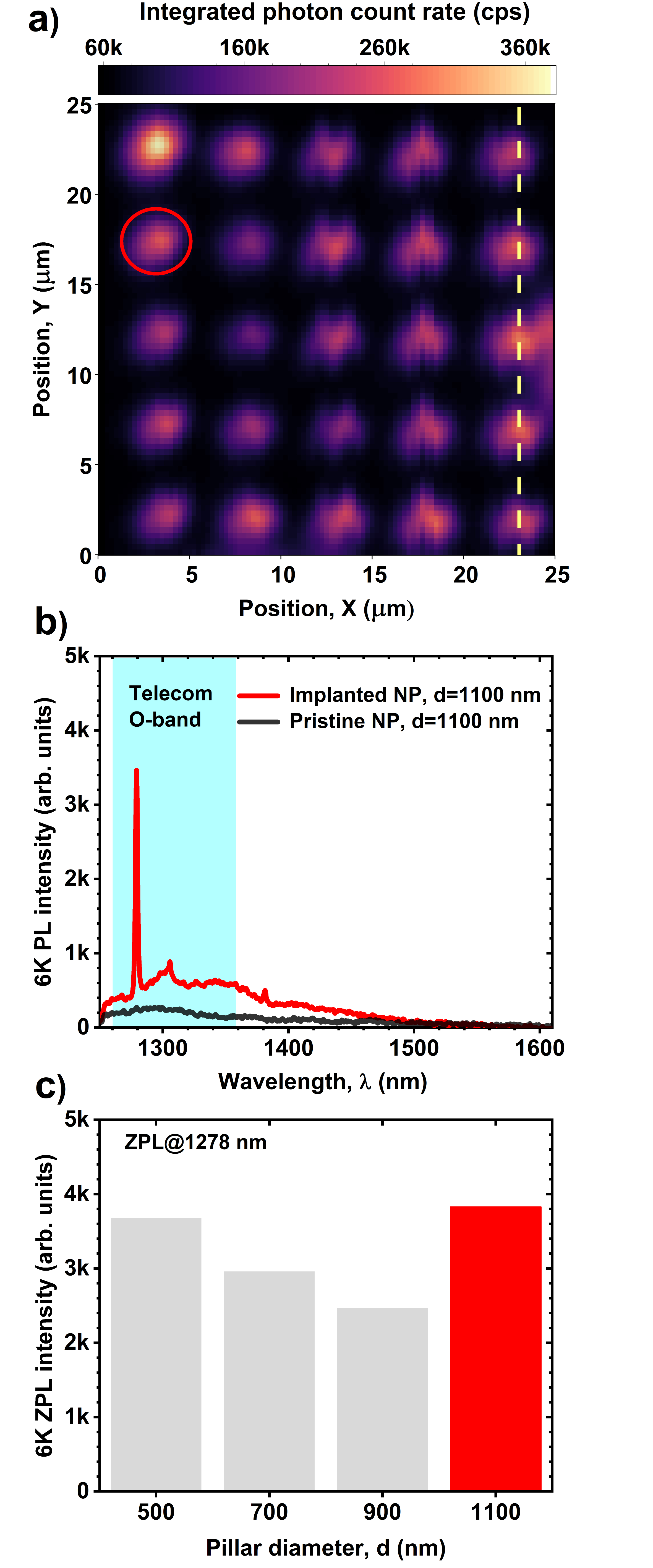}
  \caption{a) Two-dimensional confocal PL scan from one of the 5 x 5 arrays of nanopillars with a diameter of around 1100 nm. The measurement temperature is 6 K. b) Typical low-temperature PL spectrum of G-center telecom photon emitter (red line) confocally recorded on the nanopillar that is highlighted with red circle in Fig. \ref{PLspectra} a. A 6 K PL spectrum from a non-implanted nanopillar (black line) is shown for reference. c) 1278 nm ZPL intensity as a function of diameter for MACEtch fabricated pillars.}
  \label{PLspectra}
\end{figure}

Next, we spectroscopically investigate the pillars using a home-built confocal scanning microscope. A two-dimensional PL scan reveals that the arrays of Si pillars host optically active G-centers (Fig. \ref{PLspectra} a). 
A PL spectrum, recorded at a temperature of 6K, on the pillar highlighted with red circle shows PL emission in the optical telecom O-band. 
The PL spectrum (red line) exhibits a sharp zero-phonon line (ZPL) centered at 1278 nm accompanied by a broad phonon-side band at longer wavelengths (Fig. \ref{PLspectra} b), a spectroscopic fingerprint of the G-center \cite{Davies1989}. 
In contrast, no G-center related PL emission nor any other, fabrication-related, luminescent defects are detected in a pristine Si control pillar which was also patterned by MACEtch (black line in Fig. \ref{PLspectra} b). 
This indicates that the MACEtch process is advantageous for pattern transfer without introducing optically active defects in silicon. 
We also experimentally investigated the influence of the diameter of pillars on the maximum ZPL intensity for G-centers created by carbon implantation (Fig. \ref{PLspectra} c). 
An energy of 250 keV was chosen to intentionally position the generated G-centers at a depth of 600 nm, i.e. directly in the middle of the 1200 nm high MACEtched pillars. 
We find strong ZPL emission across the board in 500 nm, 700 nm, 900 nm and 1100 nm pillars.
1100 nm diameter pillars yield the best photon extraction efficiency; however, it is worthy to note that for the 500 nm diameter pillars the extraction efficiency is only slightly lower. 
Unfortunately, no PL signal was detected in the 300 nm diameter pillars, this is likely due to the high scattering losses caused by the absence of regular shaped pillars (Fig. \ref{MACEtcharray} b) which is caused by the dominance of Process I mentioned previously.

\begin{figure}
  \centering
  \includegraphics[width=0.4\textwidth]{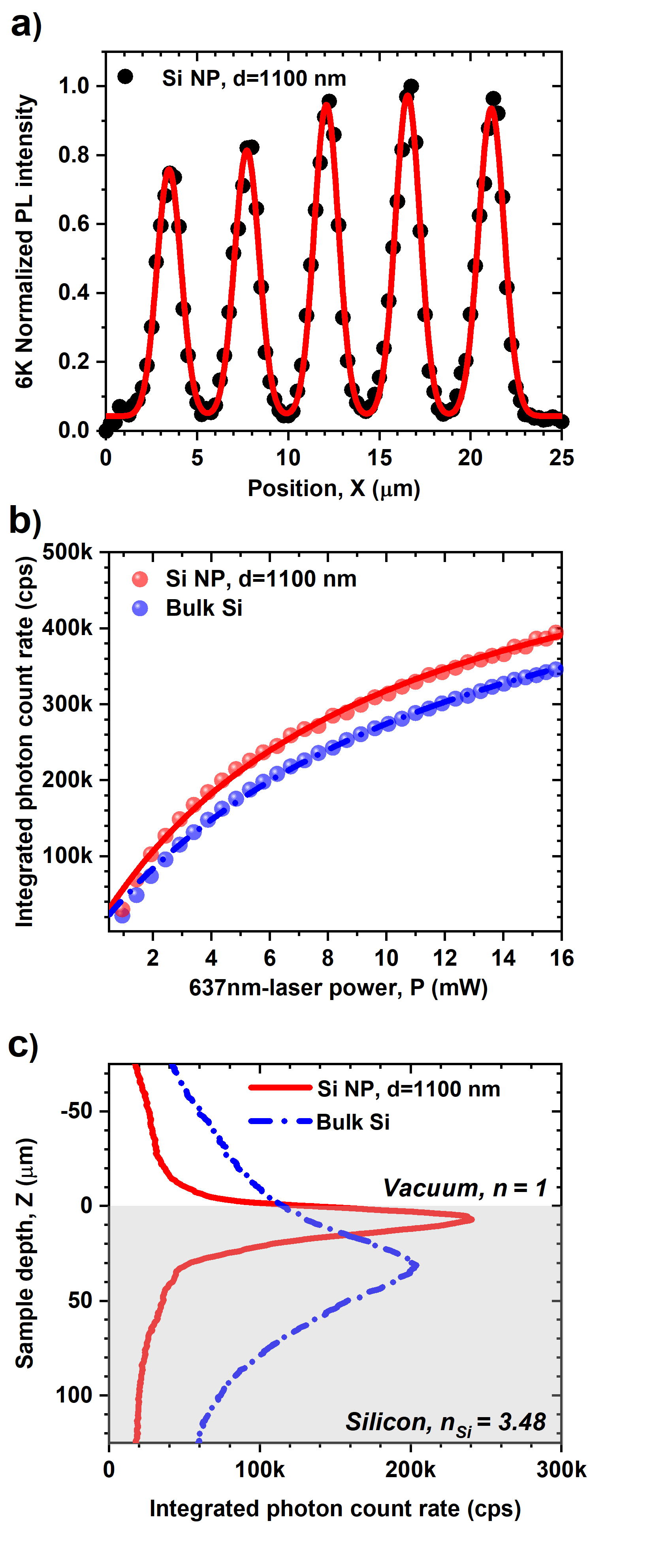}
  \caption{a) Normalized PL line scan recorded along the dashed line indicated in Fig. \ref{PLspectra} a. b) Power dependence of a telecom photon emitter integrated into an individual Si nanopillar. For reference, the power dependence from G-centers created in bulk Si, without MACEtched nanopillars, is also shown. c) in-depth (z-axis) integrated photon count rate recorded from a 1100 nm diameter pillar and non-patterned bulk Si. }
  \label{betterthanSi}
\end{figure}

A PL line scan (Fig. \ref{betterthanSi} a) along the dashed line illustrated in Fig. \ref{PLspectra} a shows strong PL emission from all pillars.
For individual pillars, we also collected a laser power-dependent saturation curve of the integrated photon count rate (Fig. \ref{betterthanSi} b). 
We obtained the background by collecting the photon count rate from the gold floor next to the pillar and subtracted from the vertically-directed G center emission. 
The maximum PL intensity was deduced by fitting the background-corrected saturation curve by the following power law:
\[
I(P)= 
\frac{I_{max}}{1+\frac{P_{0}}{P}},
\]
where \textit{I(P)} is the PL intensity as a function of the excitation power \textit{P}, \textit{P$_{0}$} denotes the saturation power and \textit{I$_{max}$} is the photon count rate when saturated. 
We estimated the saturation count rate and the saturation power to be 632 kcps and 10 mW, respectively. 
This latter leads to a power density of 1.3 MW/cm$^{2}$. 
We also found that the PL intensity of G-centers integrated into an individual silicon pillar is enhanced by a factor of 1.2 compared to the maximum emission of the G-centers in bulk Si (Fig. \ref{betterthanSi} b) because of silicon's high refractive index (\emph{n} = 3.48), which leads to a smaller collection solid angle and spatial aberration. 

Moreover, we statistically found that all nanopillars show an enhanced PL intensity as compared to bulk Si for a given laser power. 
Additionally, Fig. \ref{betterthanSi} c also shows a waveguiding effect along the pillar of the G-center emission as compared with the emission of G-centers in bulk Si for the same fluence and implantation depth. 
The waveguiding effect is evidenced by the reduction in the FWHM of the G-center emission along the pillar, which is a factor of 3.5 narrower than that of bulk Si. 
The PL noise floor in pillars also decreases by a factor of 3 compared to that of bulk Si (Fig. \ref{betterthanSi} c), which is advantageous for single-photon emitters.

Although pillars are expected to provide spatial separation of single G-center defects compared to bulk Si, the carbon fluence was found to be too high to allow detection of single G-center emitters on individual pillars with the current design of pillar diameters. 
Apart from decreasing the implantation carbon fluence, array of pillars with smaller diameters can also be fabricated at the cost of optimizing the MACEtch process. Either way would result in the creation of telecom single-photon emitters based on single G-center defects integrated into individual pillars.

\section{Conclusion}
In summary, we have demonstrated the integration of 1278 nm emitting G-centers in a two-dimensional photonic platform consisting of arrays of silicon nanopillars. We proposed a low-cost, CMOS-compatible nanofabrication process, which allows for scalability and enables the production of thousands of individual nanopillars per square millimeter, each hosting telecom photon emitters. 
We also demonstrated that metal-assisted chemical etching is an effective pathway for optically active defect-free pattern transfer in silicon, resulting in high-aspect ratio nanopillars with a 5 $\mathrm{\mu}$m-pitch that is comparable to that of integrated photonic circuits. 
We proved a vertically-directed waveguiding effect of the G-center related emission along individual pillars, which is accompanied by improved brightness, PL signal-to-noise ratio and photon extraction efficiency of the G-centers compared to that of bulk Si. We believe that this photonic platform holds great promise for hosting and scaling up single-spin color centers \cite{Kurkjian2021} and single-photon emitters \cite{Hollenbach2020,Redjem2020,Durand2021,Baron2021} in silicon coupled to individual silicon pillars. In turn, focused ion beam \cite{Krauss2017, Ohshima2018,Pavunny2021}, with its sub-micrometer writing resolution, could become the technique of choice for deterministically creating single color centers integrated into silicon photonic structures. 

\section*{Acknowledgements}
This work was financially supported by the German Research Foundation (DFG, AS310/9-1). Support from the Ion Beam Center (IBC) and clean room facility at Helmholtz-Zentrum Dresden-Rossendorf (HZDR) is gratefully acknowledged. The authors also thank Slawomir Prucnal and Shengqiang Zhou for stimulating discussions.

\section*{AUTHOR DECLARATIONS}
The authors declare no competing financial interest. 
\section*{DATA AVAILABILITY}
The data that support the findings of this study are available from the corresponding author upon reasonable request.

\end{document}